\documentclass[]{spie}
\usepackage[]{graphicx}

\title{Sternberg astronomical institute activities on site testing programs}
\author{Victor Kornilov\supit{a}
\skiplinehalf
\supit{a}Sternberg Astronomical Institute, Universitetsky pr-t, 13, Moscow, Russia
}
\authorinfo{Send correspondence to E-mail: victor@sai.msu.ru}
\begin{document}
\maketitle
\begin{abstract}
Recent Sternberg astronomical institute activities on site testing programs and technique are presented. The main attention is paid to the new modifications of MASS and DIMM data processing developed by SAI team. Four important unresolved questions affected to optical turbulence measurements are raised in the hope to be solved in nearest future.
\end{abstract}


\section{INTRODUCTION}
In spite of that SAI site testing researches have been started many years ago, this presentation responds on our activity in last five years only. First, we present a short information about finished campaign for optical turbulence measurement at Mt.~Maidanak and on-going site testing program at Mt.~Shatdzhatmaz, where the new 2.5 m telescope is planned to be install.

More detailed discussion is devoted to MASS and DIMM measurements processing, where some additional effects were taken in account in last years. Then, our plans for further enhancement of MASS and DIMM instrumentation and software are highlighted. Finally, some extra-essential questions are formulated
in hope to be solved.

Our team website is located at http://curl.sai.msu.ru/mass. A lot of documents related to the MASS and DIMM instruments and corresponding software can be found on the site. We pay main attention to the both methods details. Also, SAI Automatic Seeing Monitor (ASM) is available at  http://eagle.sai.msu.ru/ link. In the now presented works, Nicolai Shatsky, Olga Voziakova, Sergey Potanin, Boris Safonov, Matwey Kornilov took part.

\section{2005 - 2007 campaign at mount Maidanak}

Main goal of the campaign at mount Maidanak was to  finalize 1998 -- 1999 studies \cite{intas98} of the optical turbulence (OT) above the summit. Our secondary intention was to extend  our  experience in MASS observations and data processing. The campaign was performed in collaboration with Tashkent astronomical institute staff. The observations have been started in August 2005 and finished in November 2007 have covered 5 seasons.

Original MASS device installed at astrographic refractor was used as instrument. The telescope has aperture D=230~mm and focal length F=2300 mm. The original MASS \cite{MASS} was built in 2002 in the cooperation ESO + CTIO + SAI in 3 copies.  The instrument is driven by Turbina software working under GNU/Linux.

Data set was collected for 280 nights with whole length of 1022 hours. Data processing has resulted more 50\,000 OT vertical profiles. The main parameters of atmospheric OT are: free atmosphere (0.5 km and above) seeing $\beta_{free} = 0.47''$, isoplanatic angle $\theta_0 = 2.19''$. When seeing is better than its median then $\theta_0 = 2.47''$. The time constant is $\tau_0 = 3.94$ ms. Under weak turbulence $\tau_0 = 5.41$ ms. Corrected atmospheric time constant $\approx 7$ ms.

The results of this campaign are published in recent paper\cite{Maid2005} where the prospect of adaptive optics system on the Maidanak 1.5~m telescope is discussed too.

\section{2007 -- 2010 campaign at mount Shatdzhatmaz at Northern Caucasus}

\begin{figure}
\begin{center}
\begin{tabular}{cc}
\includegraphics[height=8cm]{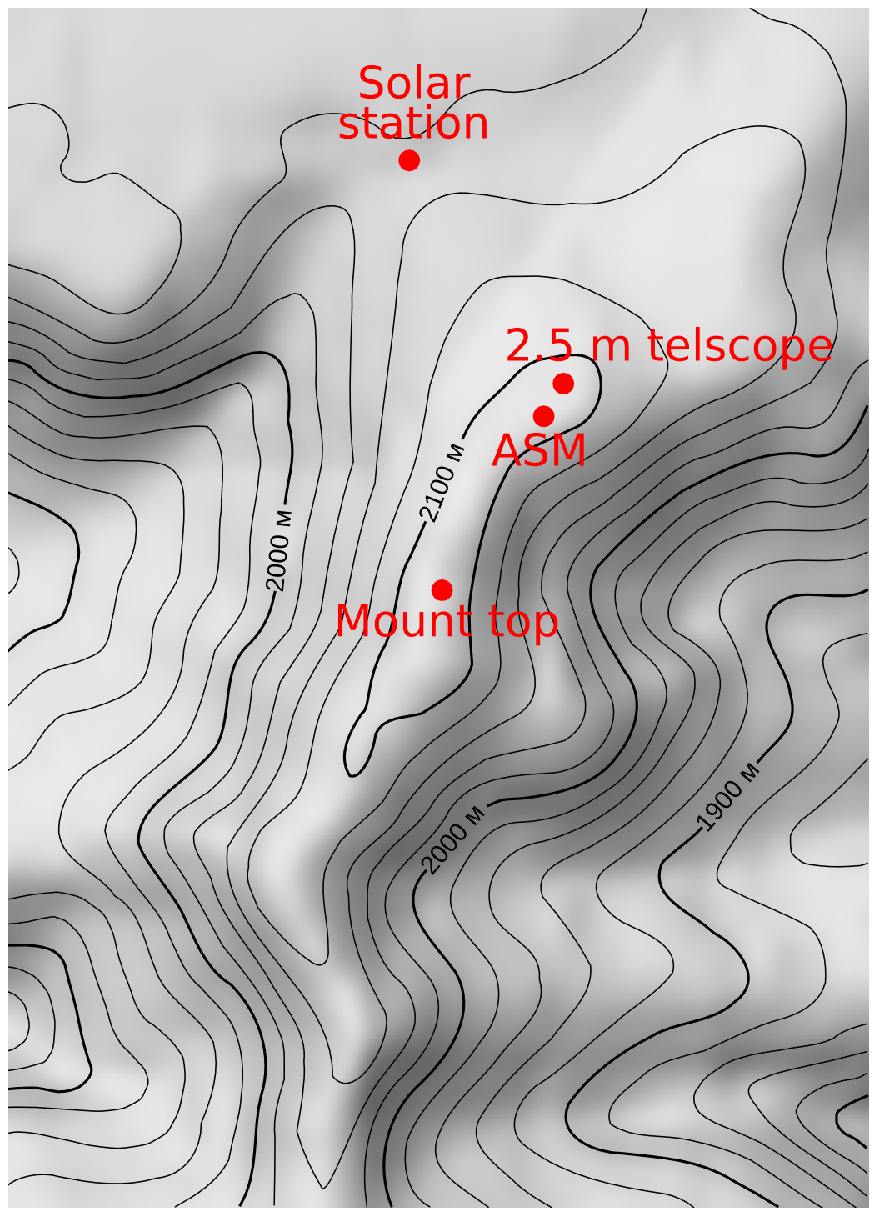} &
\includegraphics[height=8cm]{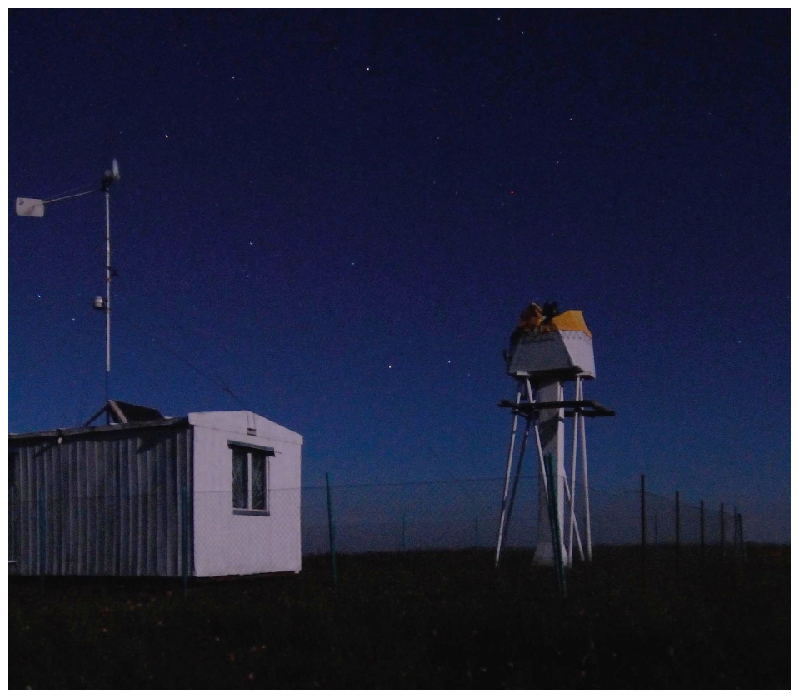}\\
\end{tabular}
\end{center}
\caption[relief] { {\it Left:} The map  2x2 km of the local relief. {\it Right:} Night time view of the ASM, July 2008.  \label{fig:relief} }
\end{figure}

Mt. Shatdzatmaz (2127 m) is located in Karachay-Cherkess Republic of Russia, 20 km southward from Kislovodsk. The mountain belongs to the Skalistiy (`Rocky') ridge which is parallel to the Main Caucasus ridge $\approx$50 km away to North. Solar station of Main astronomical observatory (Pulkovo observatory) is situated about 1 km from the place chosen for installation of new 2.5~m telescope on the top of Mt. Shatdzatmaz. The nearest part of the local relief is shown in Fig.~\ref{fig:relief} on left.

The main goal of this campaign is to collect  statistically reliable data on seeing and OT vertical distribution. In parallel, the representative information on the amount and quality of clear night sky, on atmospheric transparency, sky brightness and on-site weather parameters had to be accumulated. This information is needed to develop the optimal strategy for the 2.5~m telescope operations.

The ASM tower is installed at 40~m to SW from the spot reserved for the 2.5-m telescope.  The ASM telescope tube is raised at 6-m elevation above the ground. Seeing monitor includes next instrumental and software components:
\begin{itemize}\itemsep=-2pt
\item MASS/DIMM device to measure strength and vertical distribution of optical turbulence;
\item Telescope Meade RCX400 12'' to feed MASS/DIMM;
\item CCD finder/guider to help accurate pointing to target stars;
\item Control computer to serve MASS and DIMM data acquisition;
\item Automated enclosure of close cloth;
\item Wind direction and speed, air temperature, air humidity  sensors;
\item Boltwood clouds sensor on Solar station, self-made clouds sensor at the ASM dome;
\item Two web-cameras for internal and external overlook;
\item Controlled power supply for instruments;
\item Service computer to monitor environment and support ASM work;
\item Wi-Fi bridge to Solar station where ASM server is located;
\item Main software: {\sl mass, dimm, rcx\_scope, monitor, dome, ameba};
\end{itemize}
More detailed description of SAI ASM is available in recent paper\cite{kgo2010}. Main properties of tested site is listed below.

\begin{figure}
\begin{center}
\includegraphics[height=8cm]{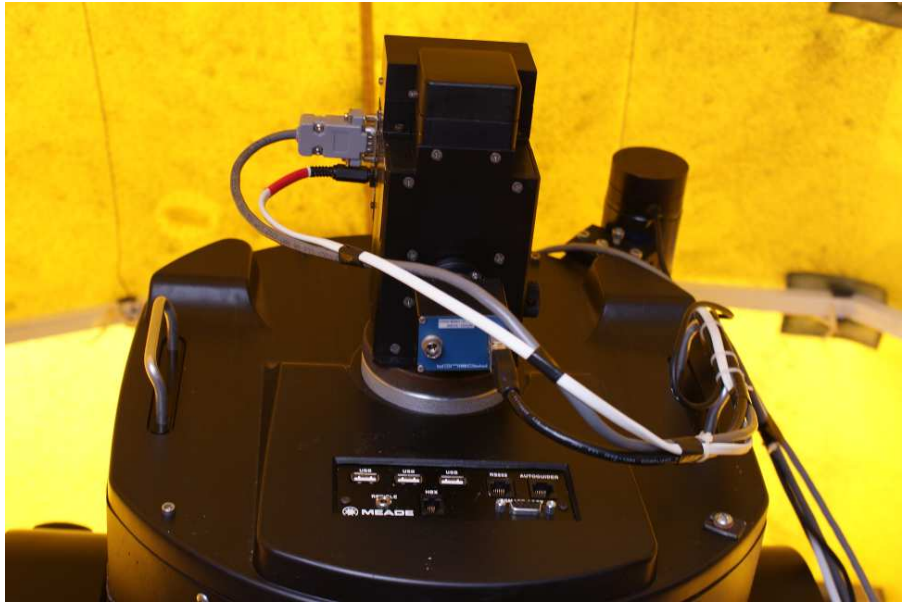}
\end{center}
\caption[example] {MASS/DIMM instrument installed on Meade telescope. The DIMM camera is Prosilica EC650 with IEEE-1394 interface. Telescope finder/guider equipped with CCD camera is seen on the right side \label{fig:example} }
\end{figure}

\subsection{Clear skies and meteo-characteristics}
An estimation of clear skies has been done on the base of sky temperature measured by IR sensor: Boltwood clouds sensor. We have defined empirically that the sky can be ``photometrically'' clear if $\Delta T = T_{sky} - T_{amb} < -22^\circ$C. For the site annual clear astronomical night skies $\approx 1340^h$  or 46\%.The maximum of the clear skies amount is observed from mid-September to mid-March, where about 70\% of the clear weather is concentrated.

Median temperature over year is $+1.8^\circ$C, temperature span is not very large: from $+15^\circ$C in summer to $-15^\circ$C -15 in winter. Median wind speed is 2.3 m/s, the dominant winds direction is from west or south-east.

\subsection{Statistics of the campaign}

The total duration of the observations $> 2700^h$ in the period  2007 November -- 2010 August. Number of accumulated profiles $> 130 000$. During the period the telescope has produced $\approx 3300$ pointings to target stars.

ASM efficiency (used time to clear sky) $\approx 50$\% in total. Over last year the efficiency $> 75$\%. In 2009 two sub-programs were added: photometric program for atmospheric extinction determination and twilight observations of the OT.

\subsection{Main characteristics of the site}

At Mt. Shatdzatmaz overall median seeing $\beta = 0.93''$. The most probable seeing value is $0.82''$. In 25\% of the time seeing is better than $0.73''$.  The free atmosphere median seeing $\beta_{free}$ is $0.51''$, its mode is $0.35''$. The best seeing (minimal OT strength) is observed in October -- November. The typical median value for that period is $\approx 0.83''$.

Median corrected atmospheric time constant is 2.58~ms, it exceeds 3.3 ms in conditions of weak turbulence. Median isoplanatic angle is $2.07''$ over whole period and $2.38''$ under condition of the seeing better than its median. Features of vertical distribution of the OT are presented in paper \cite{kgo2010}.

\section{Revision of the OT profile restoration}

{\sl Turbina} restoration algorithm is based on calculation of mean scintillation indices over {\it accumtime} (1~minute) and direct minimization of non-linear system in unknown $J^{1/2}$. The obtained cumulative distributions doesn't look physically and may be explained with joint effect of the restoration  errors and non-negativity restrictions. This effect leads to underestimated characteristic points (medians and low quartiles) of OT values in separate layers.\cite{nnls2010}

\begin{figure}
\begin{center}
\begin{tabular}{cc}
\includegraphics[height=7cm]{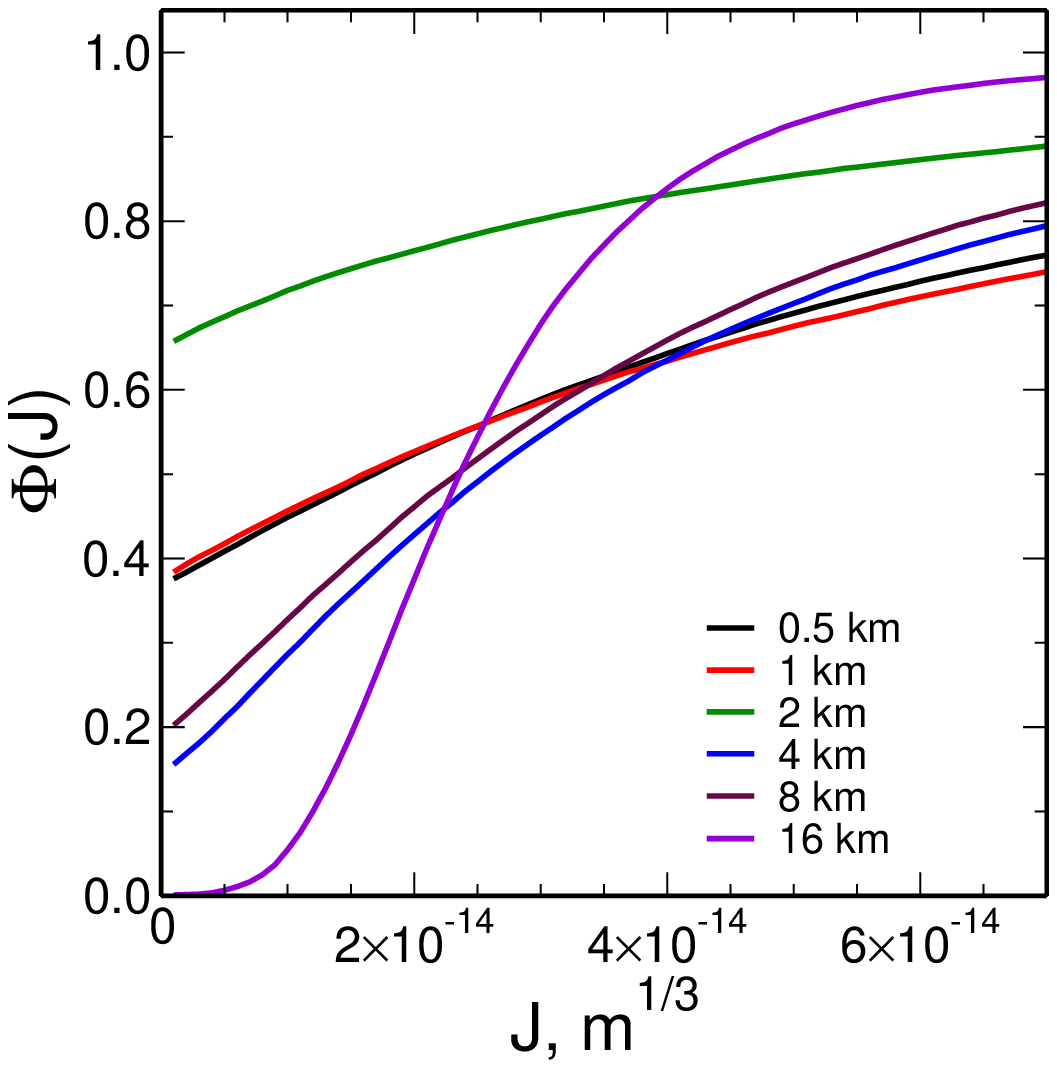} &
\includegraphics[height=7cm]{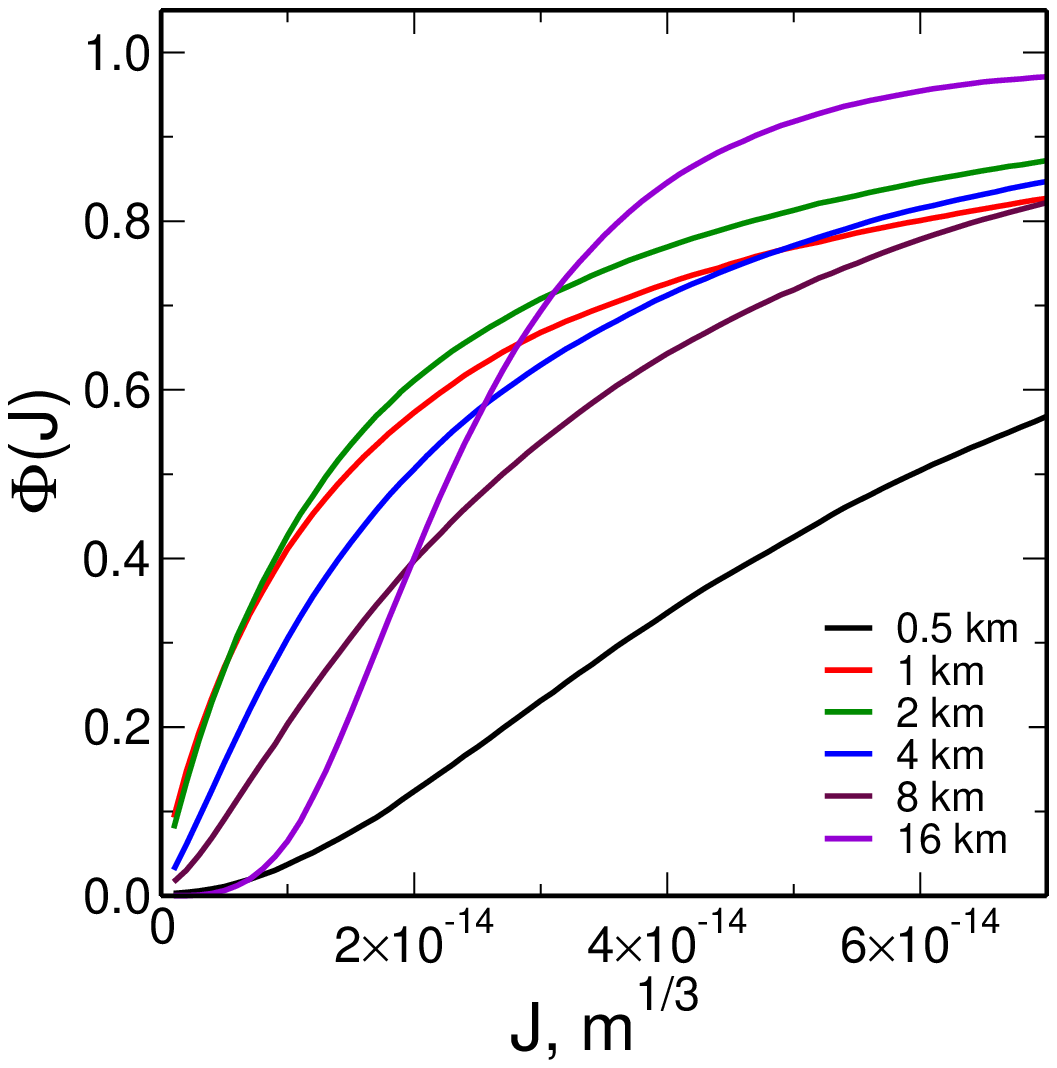} \\
\end{tabular}
\end{center}
\caption[process] {Layer-by-layer cumulative distributions of the OT above Mt. Maidanak. {\it Left:} processed with Turbina restoration algorithm {\it Right:} with help of the new algorithm \label{fig:process} }
\end{figure}

New restoration algorithm ({\sl atmos-2.96.0}) is based on Non-negative least squares (NNLS) method developed 40 years ago for the solution linear equations in terms of least squares with non-negative restrictions. New algorithm processes 1~s scintillation indices and then averages 1~s solutions. As result we have more accurate profiles and can use more layers (more than 12) in the  restoration. The cumulative distributions of the OT intensity in 6 layers after reprocessing with {\sl   atmos-2.96.0} are shown in Fig.~\ref{fig:process} on right.

Further modification of the profile restoration algorithm ({\sl atmos-2.96.7}) was done in 2010 for the processing of 2-years set of measurements at Mt. Shatdzhatmaz. This version combines both DIMM and MASS data processing. Reasons are:
\begin{itemize}\itemsep=-2pt
\item To calculate propagation effect in differential motion.
\item To restrict non-physical negative OT in ground layer.
\item To determine at once OT in ground layer.
\end{itemize}

\begin{figure}
\begin{center}
\includegraphics[height=12cm]{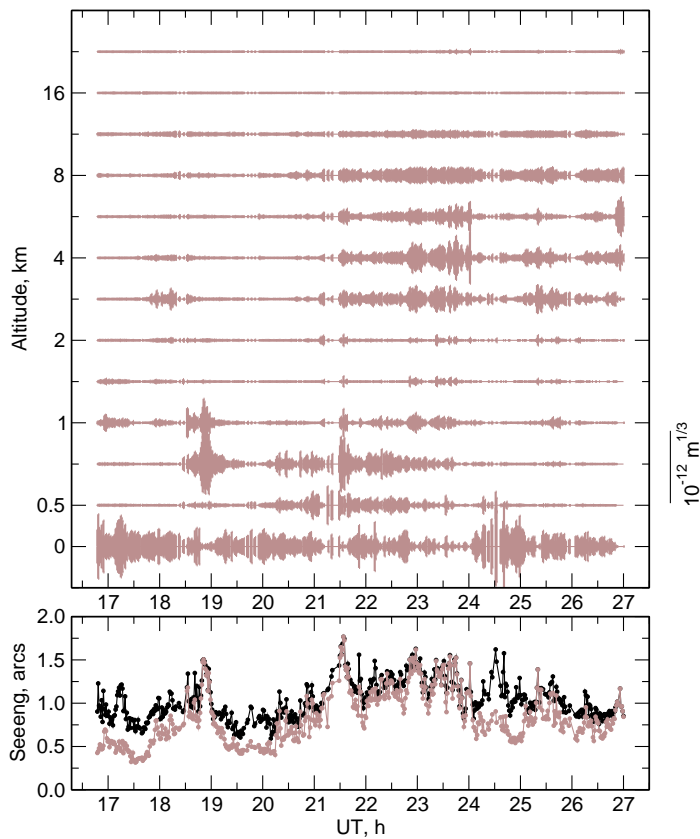}
\end{center}
\caption[night] {Evolution of the OT above Mt.~Shatdzatmaz in 2009 February 19 night as processed by
 {\sl atmos-2.96.7} version\label{fig:night} }
\end{figure}

An example of such processing for 2009 February 19 is shown in Fig.~\ref{fig:night}. For such modification we had to introduce DIMM weighting functions.

\subsection{DIMM weighting functions}

DIMM standard theory is based on near-field approximation what leads to partial loss of high turbulence power. To have an equations similar to MASS equations, the {\sl weighting functions} (WF) were introduced for the differential motion. The WFs provide next relation for transversal and longitudinal motion:
\begin{equation}
\sigma^2_{l,t} = \int C^2_n(h)\,W_{l,t}(h)\,{\rm d}h.
\end{equation}
To calculate WFs we start from earlier studies of Martin\cite{Martin87} and Tokovinin\cite{Toko2002}. For example, WF of the transversal differential motion for the case of G-tilt can be written as
\begin{equation}
W_t^{g,z}(z) = 2.403 \int_0^{\infty} {\rm d}f\,f^{-2/3} {\left[ \frac{2 J_1(\pi fD)}{\pi fD}\right]}^2 \cos^2(\pi \lambda z f^2)  \left[1 - 2\frac{J_1(2 \pi f B)}{2 \pi f B}\right].
\end{equation}

Polychromatic WFs can be computed by replacing the usual Fresnel term $\cos^2(\pi\lambda hf^2)$ with the polychromatic Fresnel filter which  is a square of the real part of Fourier transform of the incoming radiation energy distribution similarly it doing in MASS WFs calculation\cite{polichrom2003}.

\begin{figure}
\begin{center}
\begin{tabular}{cc}
\includegraphics[height=6.5cm]{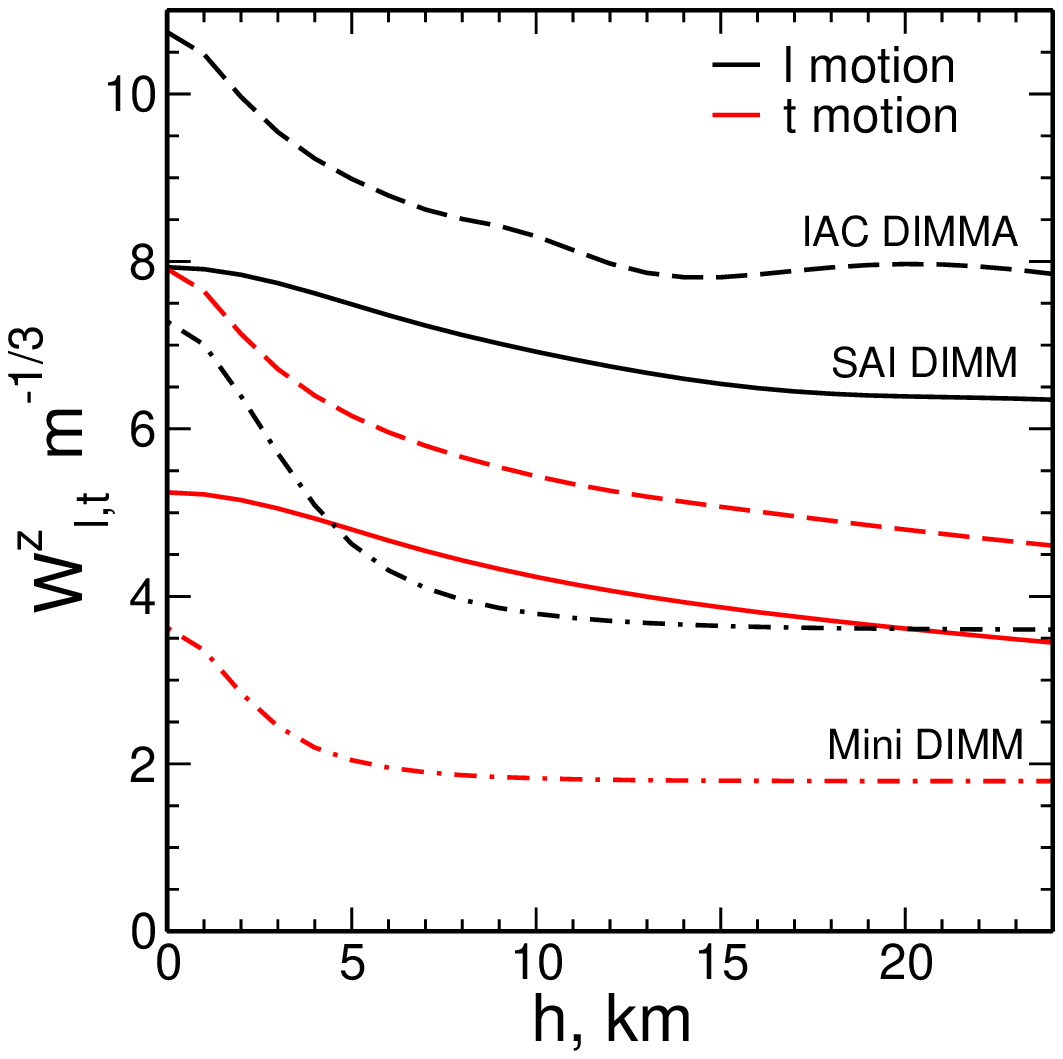} &
\includegraphics[height=6.5cm]{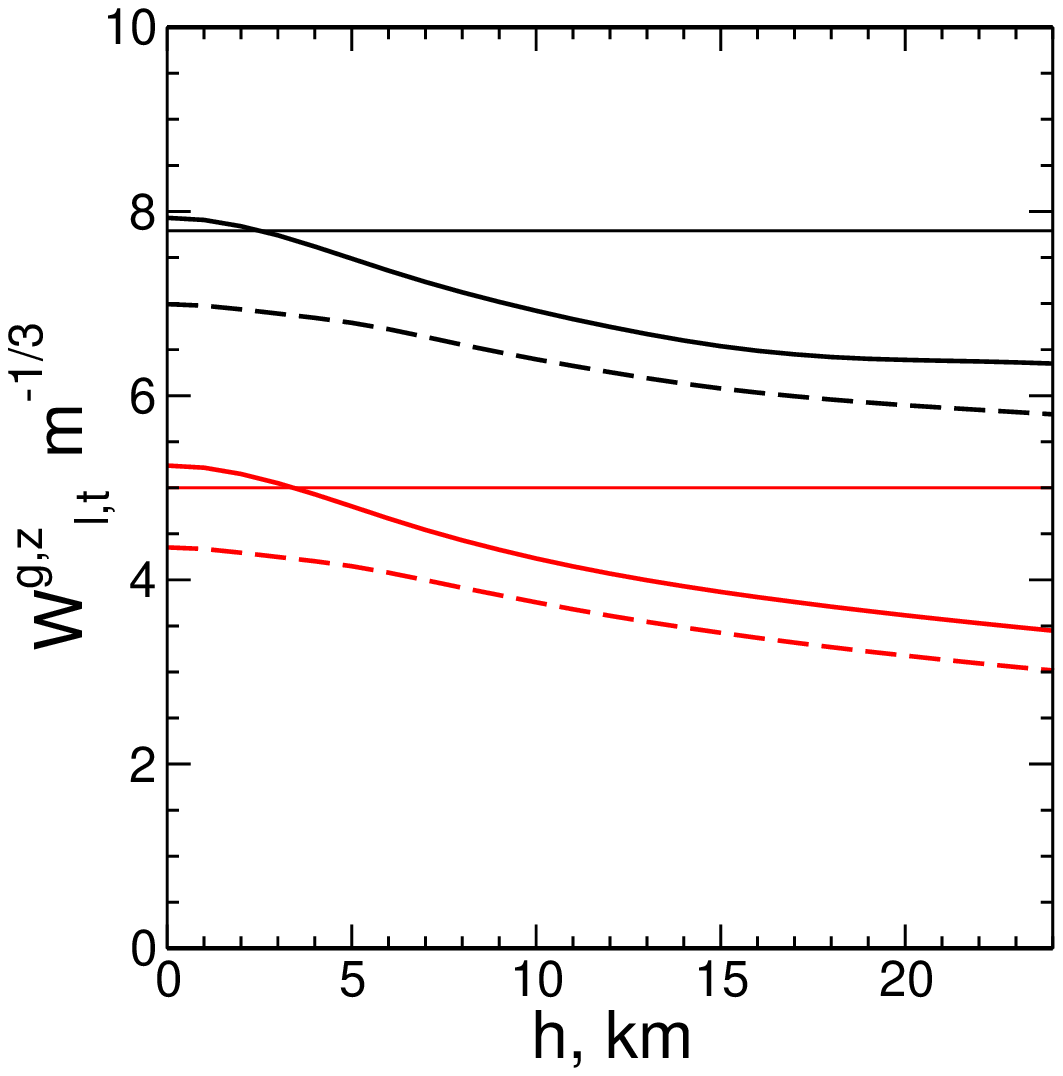} \\
\end{tabular}
\end{center}
\caption[wfs]{ {\it Left: } WFs for our DIMM (D = 0.09 m, B = 0.196 m), IAC DIMMA (D = 0.05 m, B = 0.20 m) and hypothetical mini-DIMM (D = 0.05 m, B = 0.05 m). {\it Right: } SAI DIMM WFs for Z-tilt (solid), G-tilt (dashed) and SR-tilt\cite{dimm} (thin line).   Longitudinal motion -- black, transversal -- red \label{fig:wfs} }
\end{figure}

In Fig.~\ref{fig:wfs} (on right) some WFs calculated\footnote{http://curl.sai.msu.ru/mass/download/doc/dimm\_specs.pdf} for different DIMM devices are shown. Using the WFs technique very compact DIMM device is possible to be built. Of course, parallel measurements with MASS are needed to provide OT vertical distribution. 

\subsection{DIMM finite exposure correction}

DIMM output files contain the correlation coefficient $\rho$ between adjacent measurements of the image positions. The correction is bases on its calculated dependencies on $\rho$ value. The method is approved theoretically.

For 4~ms exposure and 200~frames/s, linear approximation of the correction is in the form:
\begin{equation}
\sigma^2 = \tilde\sigma^2( 1+0.15(\rho-1)),
\end{equation}
where $\tilde\sigma^2$ is measured variance and $\sigma^2$ --- corrected

The median of measured $\rho = 0.85$ what produces the correction $\approx $2\%. Only 10\% of data requires correction greater 5\%. After December 2009 the exposure was reduced to 2.5~ms, so needed correction became  twice less.

\section{Wind in MASS data}

In MASS we see temporal fluctuation of the measured radiation mainly due to turbulence translation by the wind. In general, the variance of the flux fluctuation for the case of finite exposure $\tau$,
\begin{equation}
\sigma^2_{\tau} = \int C^2_n(h)\,W^\prime(h,\tau,w)\,{\rm d}h
\end{equation}
where $W'$ is the modified weighting function depending not only altitude $h$ but wind profile $w(h)$ and exposure.
\begin{equation}
W'(w,\tau,h) = 9.62\lambda^{-2}\int_0^\infty  {\rm d}f \, f^{-8/3} \sin^2(\pi\lambda h f^2) A(f) A_s(w,\tau,f),
\label{wdef}
\end{equation}
where $A_s$ is wind shear spectral filter having analytic expression\cite{wind2010}.

Two ultimate regimes can be emphasized: short exposure: $w\tau \ll D$ and long exposure: $w\tau \gg D$. In these cases we can take wind outside the weighting function integral. There are few problems related to this topic:
\begin{itemize}\itemsep=-2pt
\item MASS finite exposure correction
\item Atmospheric time constant estimation
\item Potential photometric accuracy evaluation
\item Wind vertical profile extraction
\end{itemize}

\subsection{Potential photometric accuracy evaluation}
\begin{figure}
\begin{center}
\begin{tabular}{cc}
\includegraphics[height=6.5cm]{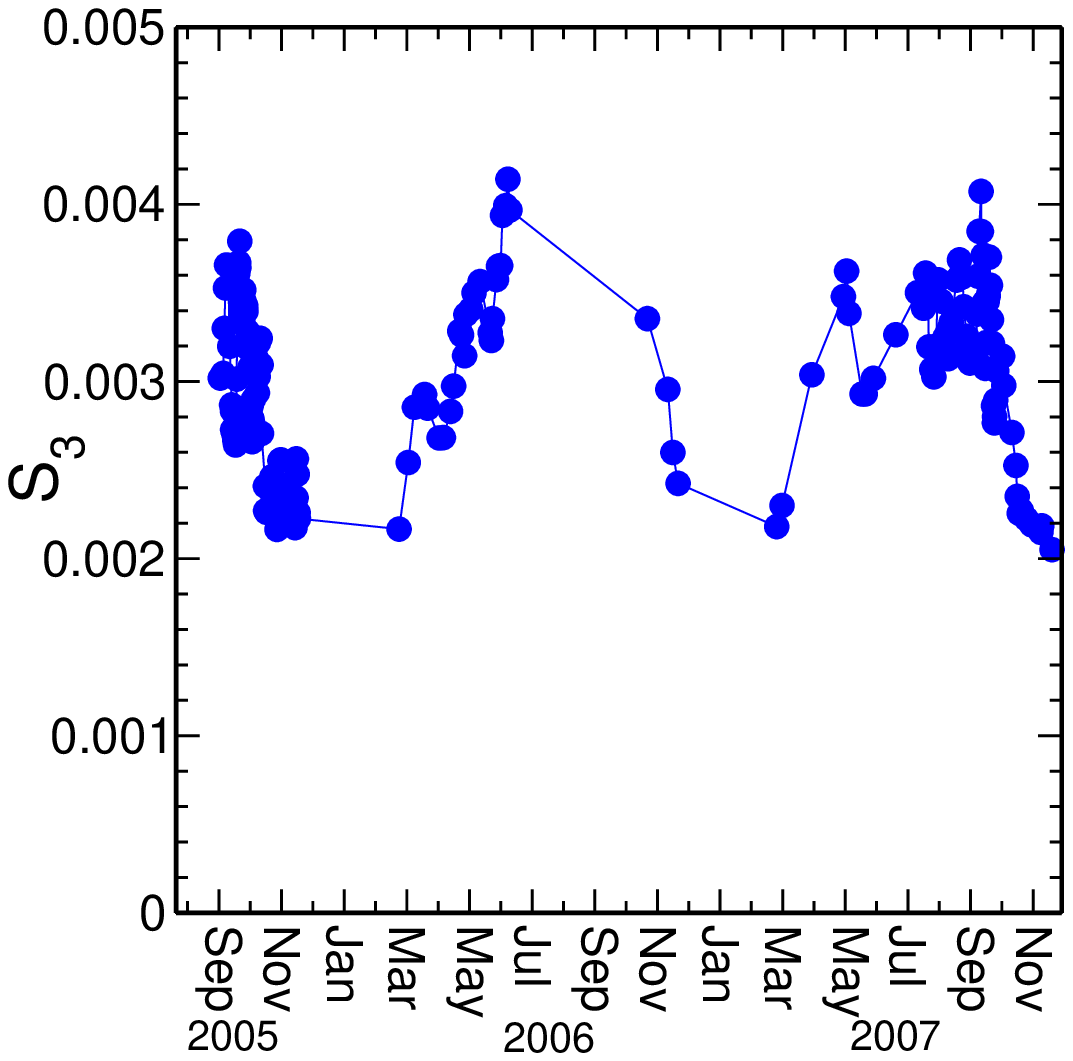} &
\includegraphics[height=6.5cm]{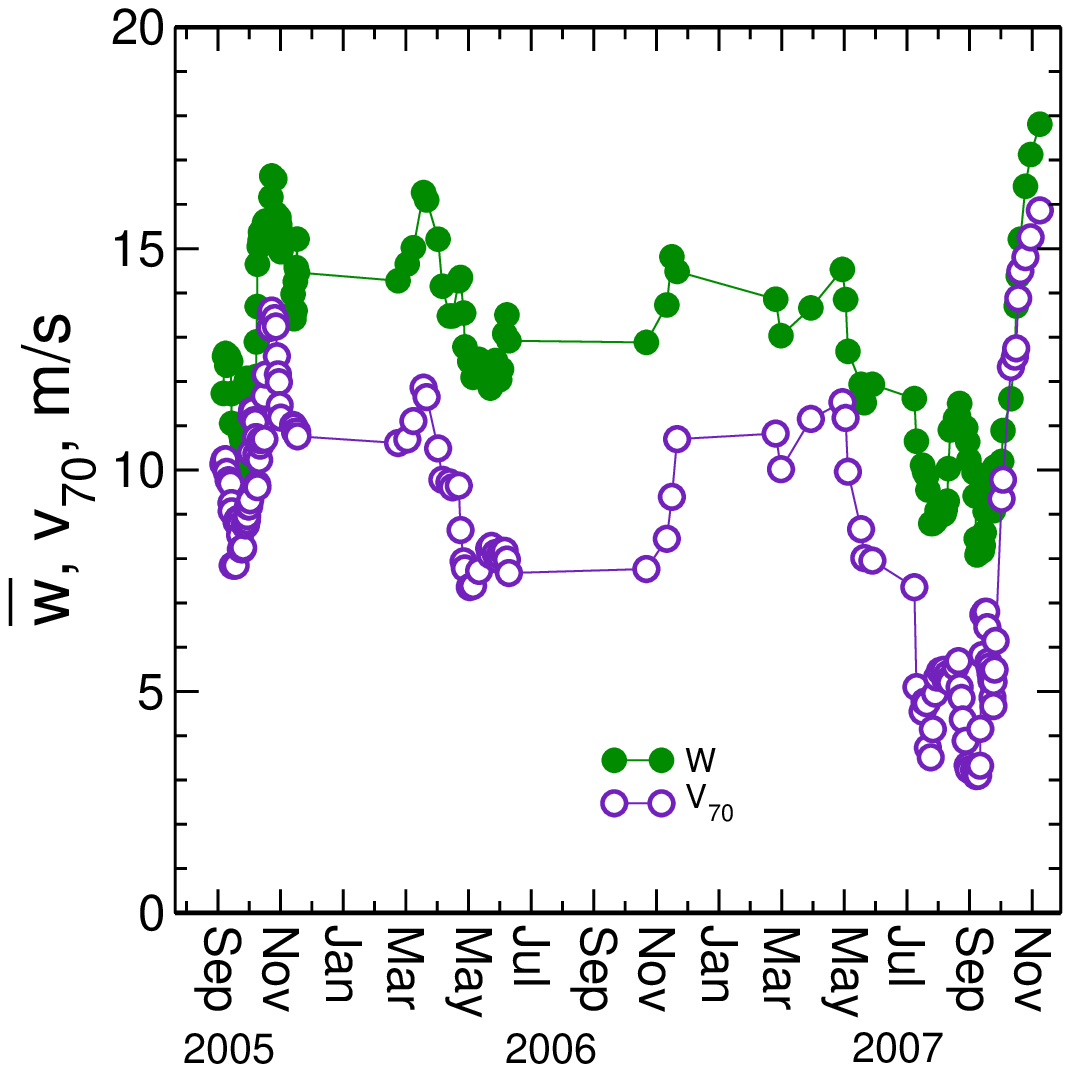}\\
\end{tabular}
\end{center}
\caption[wind-mass]{ {\it Left: } Evolution of the median values of $S_3$ index over entire period of Maidanak campaign.  {\it Right:} Comparison of the wind $\bar{\bar w}$ (filled dots) and  wind at 70~mbar $v_{70}$ (empty circles)  \label{fig:wind-mass} }
\end{figure}

In the long exposure regime ($\tau \ge 0.1$~s for MASS apertures)
\begin{equation}
\sigma^2_{\tau} = \int \frac{C^2_n(h)}{\tau w(h)}\,U^\prime(h)\,{\rm d}h
\end{equation}
From MASS data we can evaluate directly the index $S_3$ introduced in paper Kenyon et al\cite{Kenyon2006}. The index defines a photometric accuracy (scintillation noise) in aperture $D$ averaged over exposure $\tau$:
\begin{equation}
\sigma^2_{\tau} = D^{-4/3}\tau^{-1}S_3^2
\end{equation}
In the Fig.\ref{fig:wind-mass} on left, the evolution of the $S_3$ index over 2 years of the Maidanak campaign is shown.

\subsection{High-altitude wind from MASS data}
In the long exposure regime some effective wind cam be calculated:
\begin{equation}
\bar{\bar w} = \left\langle \frac{1}{w(h)}\right\rangle^{-1} = \frac{10.66\cdot M_2}{S_3^2}.
\label{eq:estim}
\end{equation}
where averaging is done with weight $C_n^2(h)\,h^2$ which is maximal at the height of $\approx 16$~km above the summit corresponding  $\approx 70$~mBar. In the Fig.\ref{fig:wind-mass} on right, the comparison of such value with data from NCEP/NCAR database is shown.

The studies give us an assurance that on the base of extended MASS data (indices for set of exposures, which are collected last 2 years) the wind profile extraction is possible. Initial approximation may be easy obtained from long exposure indices, then non-linear equation set must be solved.

\section{Some problems}

\subsection{The question No 1}
What does really DIMM measure: Z-tilt or G-tilt? The cost of this uncertainty reaches 12 -- 17\% in OT strength (see Fig.~\ref{fig:wfs} on right). When these data are used for ground layer (GL) turbulence estimation, inaccuracy as large as $\approx 30$\% can appear.

Calculations show that $\sigma^2_l/\sigma^2_t = 1.61$ for G-tilt and 1.51 for Z-tilt. Real measurements give the ratio  $\sigma^2_l/\sigma^2_t = 1.56$. High altitude turbulence can increase these ratios for both tilts. So z-tilt model is slightly preferable. Distribution of the measured  $\sigma^2_l/\sigma^2_t$ is quite wide, but corresponds an accuracy of the measurements, so data filtration by this ratio is never possible.

We need reliable experimental verification method to check any instrument and any image centering algorithm. 

\subsection{The question No 2}

Our DIMM control software ({\sl dimm}) produces the differential motion variances each {\it basetime} (1 -- 2~s). Therefore, we can analyse the motions in two timescales: faster and slower than 1-2~s. We have detected that low frequency motions have a power much greater than Kolmogorov model predicts. For low motions the median $\sigma^2_l/\sigma^2_t=1.16$.

\begin{figure}
\begin{center}
\begin{tabular}{cc}
\includegraphics[height=6.5cm]{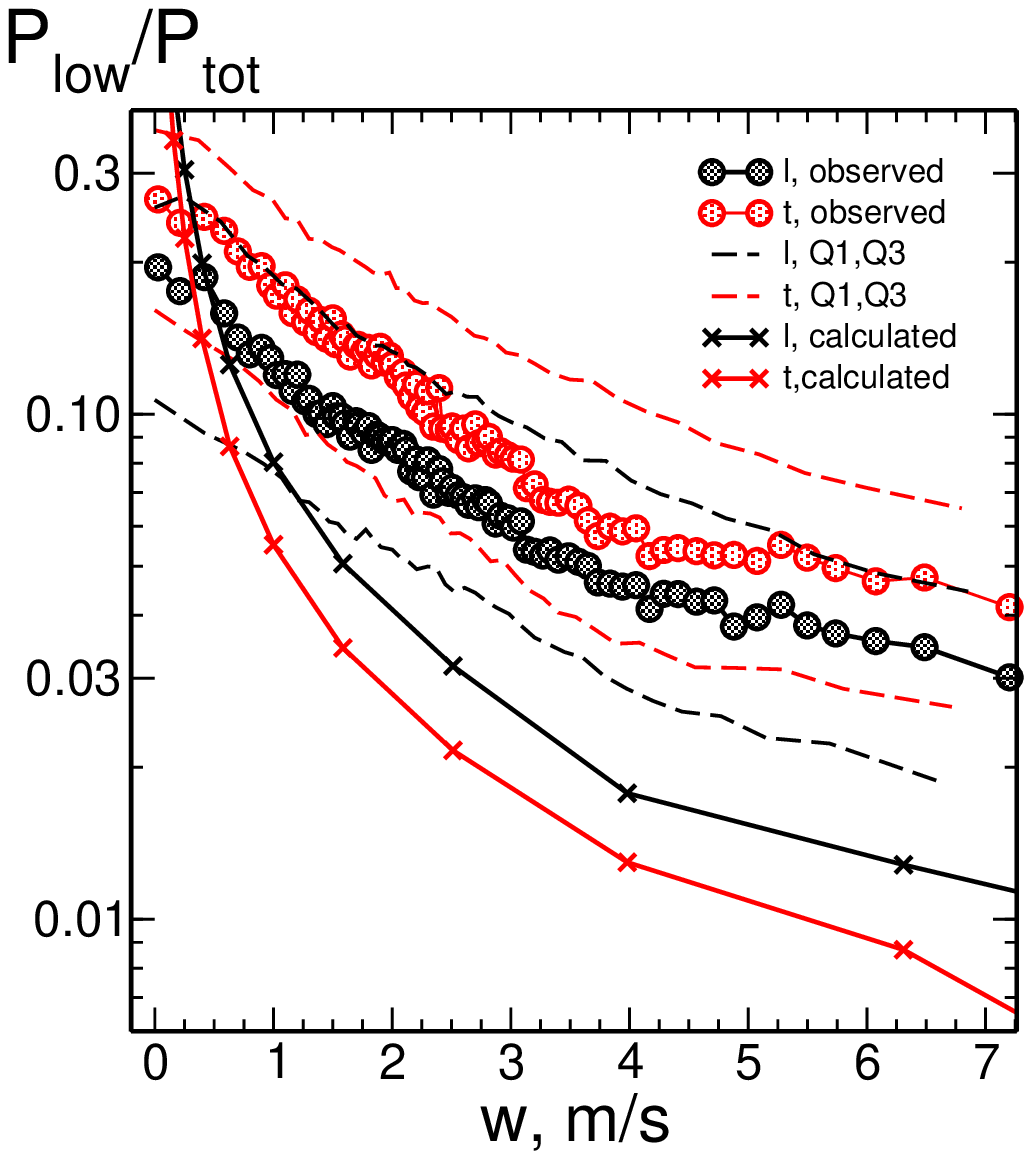}
\includegraphics[height=6.5cm]{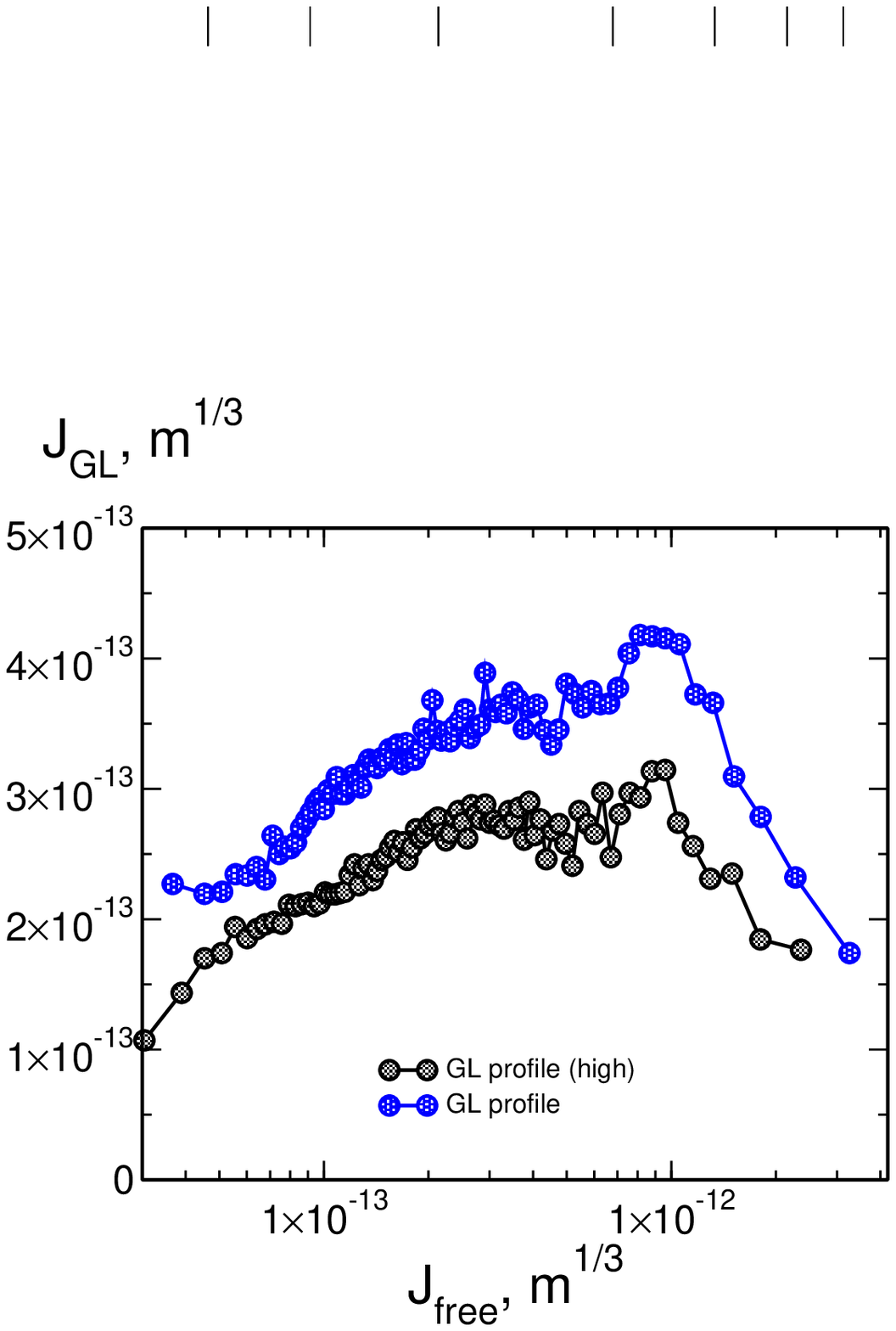}
\end{tabular}
\end{center}
\caption[low-freq]{ {\it Left:} Dependence of slow OT fraction on ground wind (circles), dashed lines are 1-st and 3-rd quartiles. Solid lines show calculated dependence in the case of Kolmogorov model. Longitudinal case is presented in black, transversal ---  in red color. {\it Right:} Ground layer turbulence versus OT in free atmosphere for two cases: full differential motion power (blue points) and high frequency differential motion (black points). \label{fig:low-ferq} }
\end{figure}

Hence, we observe the non-kolmogorov, low altitude, slow turbulence. Dependence on the ground wind is evident that is show in Fig~\ref{fig:low-ferq} on left. Is it local turbulence which isn't connected with atmosphere itself?

If we refuse this power then overall seeing decreases from $0.93''$ to $0.86''$. More important that $GL$ turbulence diminishes in 1.3 times (see Fig~\ref{fig:low-ferq} on right). Should we  include low frequency power in DIMM results?

\subsection{The question No 3}

In the last version {\sl atmos-2.96.9} the strong scintillation correction was excluded to have the strong scintillation effect undistorted. Only exact conversion of Rytov variances to scintillation indices was used. It is not enough what one can see in Fig.~\ref{fig:low-ferq} on right. The impact of strong scintillation is seen as sharp drop after $10^{-12} \mbox{ m}^{1/3}$ free atmosphere OT intensity.

The correct estimation of this effect is more important in the range  $10^{-13}\dots 10^{-12} \mbox{ m}^{1/3}$, where it is unevident. Unfortunately, there is no clear theoretical description of the effect for real astronomical conditions. Probably, an iterative processing should be needed.

\subsection{The question No 4}

Atmospheric coherence time $\tau_0$ derived from MASS data is biased. Some ways are possible to correct it: experimental calibration\cite{TMT_time}, wind profile integration, modification of  existing DESI method\cite{timeconst2002}.

It was presented already that we plan to develop the restoration of the wind profiles together with the OT profiles. As long as an accuracy of these wind profiles is unknown,  we can not estimate resulting $\tau_0$ errors. Probably such way will be impractical.

It may be shown that for short exposure regime  $w\tau \ll D | r_F$ the next approximation is valid:
\begin{equation}
\sigma^2_{\tau} = \sigma^2_0 - \frac{\tau^2}{6}\int C_n^2(h)\,w(h)^2\,U(h) \,{\rm d}h
\label{eq:sig2}
\end{equation}

The second term is the correction to zero exposure for certain index. The modification of DESI method means an usage of the full set of temporal covariances to obtain needed combination for atmospheric winds moment $w_2$.  Rejection of used empirical calibration is assumed as well.

\section{Conclusion}

The main result of our researches as in the practical use of the MASS/DIMM device and well in the processing and analysis of data of the measurement is that there is not yet  the potential of both methods  fully realized. Clearly, a simple mechanical combination of these two devices is not enough, also data processing should be common.

The posed above questions reflect the problems which need to be solved in the nearest future to provide more accurate and more robust data on optical turbulence in the earth atmosphere.

\acknowledgments 
I thank the MASS team of Sternberg Astronomical institute on whose behalf this review was been presented. I am also grateful to all conference participants for their attention to our efforts in the site testing area.


\bibliography{reference_list}   
\bibliographystyle{spiebib}   
\end{document}